\input{aipcheck}

\documentclass[
    ,final            
  ]
  {aipproc}

\layoutstyle{6x9}

\newcommand{\BEQ}{\begin{equation}}

\newcommand{\EEQ}{\end{equation}}
\newcommand{\BEA}{\begin{eqnarray}}
\newcommand{\EEA}{\end{eqnarray}}
\newcommand{\nn}{\nonumber }

\newcommand{\amsq}{{\alpha_{\rm m}^2}}

\renewcommand{\d}{{\rm d}}
\newcommand{\p}{\partial}
\newcommand{\eps}{\varepsilon}

\renewcommand{\H}{{\cal H}}

\newcommand{\N}{{\cal N}}
\newcommand{\W}{{\cal W}}
\newcommand{\Q}{{\cal Q}}

\newcommand{\taum}{{\sigma_{\rm min}}}

\newcommand{\half}{\frac{1}{2}}
\def\dbarrm {\,\,{\mathchar'26\mkern-9mu{\rm d}}}                       %
\def\hbarit {{\mathchar'26\mkern-7muh}}                             %

\begin{document}

\title{Quantum Brownian motion and \\ its conflict with 
the second law}

\author{Theo M. Nieuwenhuizen}
{address={Institute for Theoretical Physics, 
Valckenierstraat 65, 1018 XE Amsterdam, The Netherlands}}

\author{Armen E. Allahverdyan}
{address={Yerevan Physics Institute,
Alikhanian Brothers St. 2, Yerevan 375036, Armenia }}
\date{\today}

\begin{abstract}

The Brownian motion of a harmonically bound quantum particle 
and coupled to a harmonic quantum bath is exactly solvable. 
At low enough temperatures the stationary state is non-Gibbsian 
due to an entanglement with the bath. 
This happens when a cloud of bath modes
around the particle is formed.
Equilibrium thermodynamics for particle plus bath together, 
does not imply standard thermodynamics for the particle itself at low $T$.
Various formulations of the second law are then invalid. 
First, the Clausius inequality can be violated. 
Second, when the width of the confining potential is suddenly changed, 
there occurs a relaxation to equilibrium during which
the rate of entropy production is partly negative. 
Third, for non-adiabatic changes of system parameters the rate of 
energy dissipation can be negative, and, out of equilibrium, cyclic
processes are possible which extract work from the bath.
Conditions are put forward under which perpetuum mobile of the 
second kind, having several work extraction cycles, 
enter the realm of condensed matter physics.

 \end{abstract}

\maketitle


{\bf Introduction.}
There are not two fundamental theories of nature, quantum mechanics and 
thermodynamics. There is only one: quantum mechanics, while
thermodynamics must emerge from it.
The universal character of equilibrium thermodynamics led to the 
general expectation that in one way or another, thermodynamics
will apply to the full quantum domain~\cite{landau}. 
Few people have taken the painful road to check this emergence,
yet this is what we have set out to do. Here we discuss the results
for quantum Brownian motion that have been presented
~\cite{ANQBMprl,NAlinw} and were discussed in the scientific literature
\cite{AIP}.

Brownian motion has numerous applications in condensed matter physics
\cite{klim,weiss,risken,leggett}, atomic physics
\cite{klim}, quantum optics and chemistry \cite{gardiner}. 
Some realizations involve weak coupling with the thermal 
bath \cite{gardiner}. However, there are well-known experimental 
situations, which are essentially far from the weak-coupling regime.
Here standard thermodynamics may not apply.
The main example of this is the case of weak links between superconductive
regions,  the so-called Josephson junctions, in their overdamped regime 
\cite{van,likho}, where the relevant ranges of parameters 
were achieved already twenty years ago. Even in quantum optics, which has
often been satisfactorily described by weak-coupling  theories 
\cite{gardiner}, there are recent experiments showing the 
necessity for moderate and strong coupling approaches~ \cite{oe}.

{\bf The Hamiltonian.} We consider an `ideal gas' of non-interacting
harmonic oscillators coupled to a bath.
For the total Hamiltonian $\H_{\rm tot}=\H+\H_B+\H_I$ 
we thus assume~\cite{weiss}
\BEA
\label{hamiltonian}
\H=\frac{p^2}{2m}+\half a x^2, \quad \H_B=\sum_{i}\left [
\frac{p_i^2}{2m_i}+\frac{m_i\omega_i^2}{2}x_i^2\right], \quad 
\H_I=\sum_i\left[-c_ix_ix+\frac{c_i^2}{2m_i\omega_i^2}\,x^2\right ],
\EEA
with $\H$ describing the particle, $\H_B$ the bath, 
and $\H_I$ the interaction.

The bath is assumed to have uniformly spaced modes 
$ \omega_i=i\Delta,\qquad i=1,2,3,\cdots$.
The thermodynamic limit for the bath is taken  by sending 
$\Delta\to 0$, which induces relaxational behavior.
For the couplings we choose the quasi-Ohmic Drude-Ullersma 
spectrum, where \cite{weiss}
$J(\omega)=\half{\pi}\sum_i ({c_i^2}/{m_i\omega_i})
\delta(\omega-\omega_i)=
{\gamma \omega\Gamma^2}/(\omega^2+\Gamma^2)$,
where $\gamma$ is the damping constant. 
We shall assume that the Debye cutoff $\Gamma$ is large.

We describe an ensemble of closed total systems with conserved energy, 
except for the periods when work is done on it by externally changing 
$m$ or $a$. We take as initial density matrix the Gibbs distribution
$\exp(-\beta \H_{\rm tot})/Z_{\rm tot}$, with $Z_{\rm tot}$ the
partition sum.

{\bf Equilibrium state.}
In Gibbsian equilibrium of the total system the
free energy reads
$F_{\rm tot}(T,\gamma)= F_B(T,\gamma=0)
+F_p(a,\gamma,\Gamma,m,T)$.
The free energy of the bath itself,
\BEA\label{Ftotal2=} F_B(T,\gamma=0)=\frac{T}{\Delta}
 \int_0^\infty\d\omega\ln(1-e^{-\beta\,\hbarit\omega})
= -\frac{\pi^2T^2}{6\,\hbarit\Delta}, \label{comrad}\EEA
is of order $1/\Delta$, showing the extensivity of the bath.
The Brownian particle adds to this 
\BEA
\label{artush}
 F_p=T\left[\ln\Gamma\left(\frac{\beta\,\hbarit\Gamma}{2\pi}\right)
-\ln\Gamma\left(\frac{\beta\,\hbarit \omega_1}{2\pi}\right)
-\ln\Gamma\left(\frac{\beta\,\hbarit \omega_2}{2\pi}\right)-
\ln\Gamma\left(\frac{\beta\,\hbarit \omega_3}{2\pi}\right)
-\ln\frac{\beta\,\hbarit\omega_0}{(2\pi)^2}\right].
\EEA
It contains three characteristic, temperature independent of frequencies.
For small $\gamma $ (underdamping) they read:
$\omega _{1,2}= \pm i \omega _0 +{\gamma}/{2m}$,
$\omega _{3}=\Gamma  - {\gamma}/{m}$.
On the other hand, for strong damping $\gamma^2\gg am$,
$\omega _{1}={a}/{\gamma}$,
$\omega_2=({\gamma}/{m})(1-{am}/{\gamma^2})$,
$\omega _{3}= \Gamma-{\gamma}/{m}$.

{\it Effective temperatures.}
We shall now study two objects,  $T_x=a\langle x^2\rangle$
and $ T_p=\langle p^2\rangle/m$,  that would in classical
equilibrium be equal to $T$ and which we shall interpret below
as effective temperatures. 
As in the classical situation, it holds that
$T_x=2a{\partial F_p}/{\partial a}$, $T_p
=-2m{\partial F_p}/{\partial m}$.
For $\gamma\to 0$ one gets the weak-coupling result known from
textbooks, 
$U\equiv\half T_x+\half T_p=T_x=T_p={\half\,\hbarit \omega _0}
\coth\half{\beta\,\hbarit\omega _0}$, with  $\omega_0=\sqrt{a/m}$. 
At large $T$ one gets
\BEA\label{TxTplargeT}
T_x=T+\frac{\beta\,\hbarit^2a}{12m},\quad
T_p=T+\frac{\beta\,\hbarit^2(a+\gamma\,\Gamma)}{12m},\quad
F_p= T\ln\beta\,\hbarit\sqrt{\frac{a}{m}}+
\frac{\beta\,\hbarit^2(a+\gamma\,\Gamma)}{24m}.
\EEA
At low $T$ and for strong damping one has
\BEA
T_x=\frac{\,\hbarit a}{\pi\gamma}\ln\frac{\gamma^2}{am}
,\quad
T_p=\frac{\,\hbarit \gamma }{\pi m}\ln\frac{\Gamma m}{\gamma}+
\frac{\,\hbarit a}{\pi \gamma} 
,\quad
F_p=\frac{\,\hbarit \gamma}{2\pi m}
\ln\frac{e\Gamma m}{\gamma}+
\frac{\,\hbarit a}{2\pi \gamma}\ln\frac{\gamma^2}{a m}.
\label{mega3}
\EEA
The fact that $T_p\ge T_x>0$ at $T=0$ is related to the quantum
nature of the problem. 

{\bf Generalized thermodynamic formulation.}
The Wigner function has a quasi-Gibbsian expression, 
since there occur two  temperature-like variables,
\BEQ \label{Wpx=}
W(p,x)=W_p(p)W_x(x)=\frac{e^{-p^2/2mT_p}}{\sqrt{2\pi mT_p}}\,\,
\frac{e^{-ax^2/2T_x}}{\sqrt{2\pi T_x/a}}.\EEQ
The Boltzmann entropy of momentum and coordinate is
\BEA 
\label{kow3}
&&S_p=-\int \d p W_p(p)\ln[W_p(p)\,\sqrt{\,\hbarit}\,]=
\half \ln \frac{emT_p}{\,\hbarit},
\quad S_x= \half \ln \frac{eT_x}{\,\hbarit a},
 \EEA
while the von Neumann entropy reads~\cite{weiss}
$S_{vN} =(v+\half)\ln(v+\half) -(v-\half)\ln(v-\half)$, with
$v={\Delta p\,\Delta x}/{\,\hbarit}=
\sqrt{{\langle p^2\rangle\langle x^2\rangle}}/{\,\hbarit}=
\sqrt{{mT_pT_x}}/\,\hbarit \sqrt{a}$.
The first terms in its large $v$-expansion,
$S_{vN} =\ln v+1$, coincide with the total Boltzmann entropy
$S_B=S_p+S_x$.

{\it Internal energy and interaction energy.}
The energy of the central particle reads
$U=\langle\H\rangle=\half T_p+\half T_x$.
The energy of the cloud of bath modes 
that surround the particle is   
$ U_{\rm int}=U_{\rm tot}-U_B(\gamma=0)-U=U_p-U=
\Gamma {\p F_p}/{\p\Gamma}$.
At high temperatures one has 
$ U=T+{\beta\,\hbarit^2}(2a+\gamma\,\Gamma)/24m$, 
$U_{\rm int}={\beta\,\hbarit^2\gamma\,\Gamma}/{24m}$.
Since the energy of the cloud involves $\,\hbarit$,  it is a quantum effect.
At low temperatures one has, if $\Gamma$ is large,
$U_{\rm int}={\,\hbarit\gamma}/{2\pi m}+{\cal O}(T^4)$.

{\it Implementation of the first law.}
Given the Hamiltonian $\H(p,x)$ of the subsystem, its
energy is $U=\langle \H\rangle=\int \d p\d x \H(p,x)W(p,x)$,
where $W$ is the Wigner function of the subsystem.
When we change a system parameter, the energy changes as
$\d U = \d\int \H W = \int\H \, \d W+\int W\, \d \H
\equiv \dbarrm {\cal Q}+\dbarrm {\cal W}$. 
The first term  represents the
variation due to the statistical redistribution of the 
phase space, which we identify with the change in heat 
$\,\dbarrm {\cal Q}$. 
The last term results from the change in the Hamiltonian,
so it is a mechanical, non-statistical object, which
we associate with the work $\,\dbarrm {\cal W}$ 
done by external sources.

{\it Generalized free energy and the second law.}
The definition of the effective temperatures 
admits a thermodynamical interpretation.
The free energy $F$ for a two-temperature
system is  defined as ~\cite{1} 
$F=U-T_pS_p-T_xS_x=-\half T_p\ln m T_p-\half T_x\ln({T_x}/{a})$.
For adiabatic changes in $m$ or $a$ one has 
$\d F=- S_x\d T_x- S_p\d T_p+\dbarrm {\cal W}$,
with work
$\dbarrm {\cal W}
=\langle{\partial \H}/{\partial m}\rangle\d m
+\langle{\partial \H}/{\partial a}\rangle\d a
=-T_p{\d m}/{2m}+T_x{\d a}/{2a}$.
Due to the first law this yields the 
second law for situations with two temperatures,
$\dbarrm {\cal Q}= T_p\d S_p+T_x\d S_x$,
 in close analogy with those proposed recently for 
nonequilibrium glassy systems \cite{1}
and black holes~\cite{Nbh}. 
Notice that $F$ pertains to the particle alone, and does not
satisfy Gibbsian thermodynamics,
while the Gibbsian $F_p$ of Eq. (\ref{artush}),
 relates to the whole equilibrium system, i.e., 
to the particle and the cloud of bath modes around it. 
There are many physical systems, such as a Josephson
junction strongly coupled to the electromagnetic field, 
where the natural object to study is $F$,  
relating to properties of the junction only.

{\it Violation of the Clausius inequality at low $T$.}
The Clausius inequality $\dbarrm {\cal Q}\le T\d S_{vN}$ is one of the
formulations of the second law. 
At $T=0$ it says that no heat can be taken from the bath, 
at best heat can go from the subsystem  to the  bath. 

When we change $a\to a+\d a$, $\dbarrm {\cal Q}$ is of order $-T^2\d a$, 
while $T\d S_{vN}$ is of order $-T\d a$. 
In the case $\d a >0$, where an amount of work 
$\dbarrm\, {\cal W}_{\rm rev}\sim \d a$ is done on the system, 
the Clausius relation is thus violated at low $T$.
Likewise one can consider the variation of the (effective) mass $m$.
Here one has $\dbarrm {\cal Q}
=\,\hbarit \gamma\,\d m/{2\pi m^2}$.
Now there is a transfer of heat even
if the bath temperature is zero. Thus, violation of the Clausius inequality 
is even stronger in this case.
This situation with $\dbarrm\, {\cal W}_{\rm rev}<0$ corresponds to the work 
performed by the system on the environment. That the heat 
comes from the cloud of bath modes, is confirmed by the fact that 
$ \dbarrm {\cal Q}_{\rm rev}=-\d U_{\rm int}$.

{\it Violation of the Landauer bound for information erasure.}
A further aspect is the squeezing of phase space and entropy, relevant 
for computing in the quantum regime. We have shown that the so-called 
Landauer bound $-\dbarrm \Q\ge k T\,\ln 2$ for the minimal energy 
dispersion when erasing one bit of information is violated in a similar 
manner~\cite{ANQBMprl}. This connection arises because the Landauer 
bound is just based on the Clausius inequality.

{\bf Dynamics from a non-equilibrium state.}
{\it  Energy oscillation at low $T$.} 
We consider the dynamical evolution of a system initially in 
equilibrium characterized by a spring constant $a_0$, which
at $t=0$ is instantaneously changed to $a_1=a$. 
These parameters are connected as $ a_0=(1-\alpha_0)a$ and 
we assume that $|\alpha_0|\ll 1$. 
For strong damping one has for the energy of the subsystem
$U(t)=U(\infty)+(\,\hbarit a\alpha_0/2\pi\gamma)f_1(at/\gamma)$.
The behavior of $f_1(\sigma)$ at different temperatures is presented in
Fig. 1a. For $\alpha_0>0$ it says that, after initially energy has been put
on the particle by the change of $a_0\to a>a_0$, this energy leaks away 
into the bath. At low $T$, however, too much leaks away, 
and a part has to come back.
This ``bouncing'' is familiar of the noise correlator, which behaves 
like $f_1$.

\begin{figure}
\includegraphics[width=6cm]{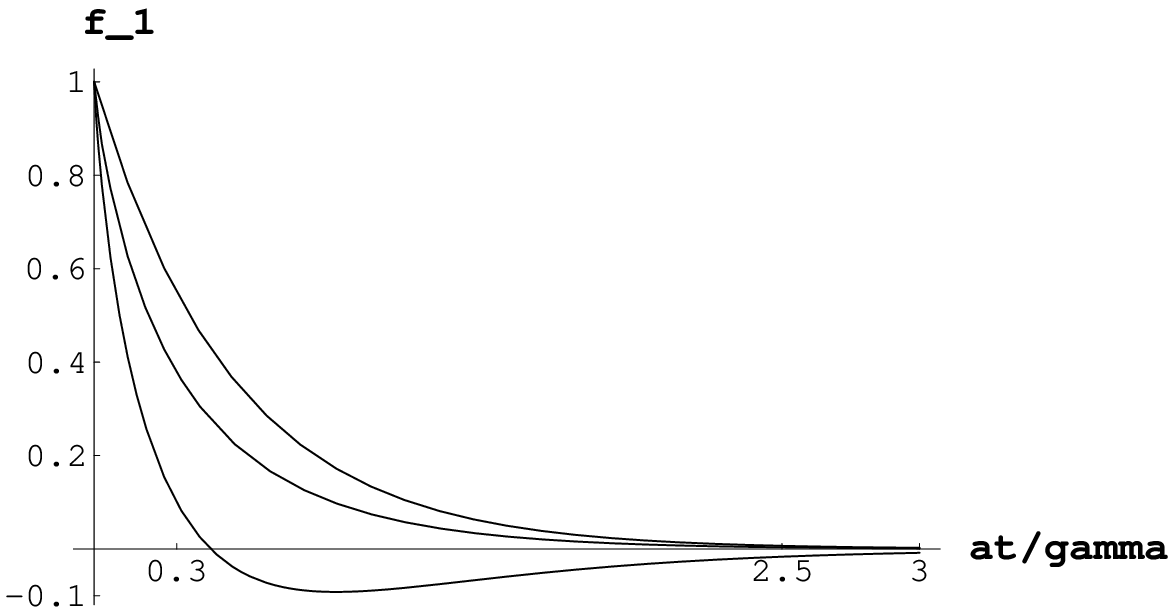}\hspace{1cm}
\includegraphics[width=6cm]{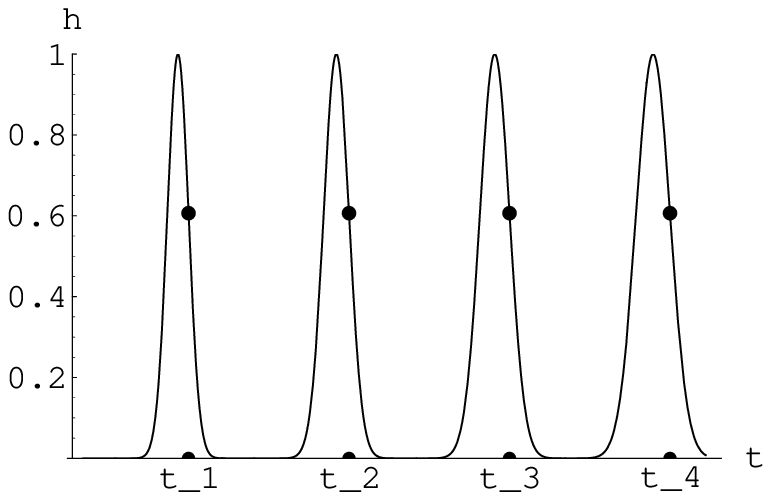}
\caption{a) Normalized excess energy $f_1$ as a function 
of $at/\gamma$, 
for different values of the temperature. Upper curve: $T\to\infty$.
Middle curve: $T=\,\hbarit a/2\gamma$. 
Lower curve: at $T=0$ it is non-monotonous. \hspace{1cm}
b) Schematic plot of the cyclic changes in the spring constant,
where successive cycles are slower and slower.
$h$ characterizes the size of the change and $t$ denotes
the dimensionless time. The interval $-\infty<t<t_1$
marks the process that establishes the nonequilibrium state at
$t=t_1$. The picture shows three full cycles, in the intervals
$t_i<t<t_{i+1}$, ($i=1,2,3$). Their start and end points are indicated by
bullets.}
\label{fig1}
\end{figure}

{\it Entropy production.}
The rate of production of Boltzmann entropy is found after a lengthy
identification of its flux. The result can
be negative at moderate $T$~\cite{NAlinw}.
For $T=0$ and strong damping one has 
with $\sigma=at/\gamma$ and  $\sigma_{min}=0.879$:  
${\d _iS_x}/{\d t}\sim + \alpha_0^2(\sigma-\taum)(\sigma-\taum 
+\eps)$, where $\eps=am/\gamma^2$.
This is negative for $\taum-\eps<\sigma<\taum$, 
or $\Delta t=m/\gamma$, the characteristic timescale of the momentum.

{\it Smooth changes of the spring constant.}
Let us now consider the case where, 
starting from the equilibrium state $a(-\infty)=a$, the spring constant 
$a(t)=[1-\alpha(t)]a$, is slightly changed ($|\alpha(t)|\ll 1$) in a 
smooth way.
The rate of work added to the system is
${\d \W}/{\d t}={\d \W_{\rm rev}}/{\d t}+{\d \Pi}/{\d t}$,
with the adiabatic (`reversible') and dispersed contributions
\BEQ \frac{\d \W_{\rm rev}}{\d t}=
-\frac{\gamma\,\alpha'(\tau)}{4m}\,
\left[T_x+\frac{\,\hbarit a}{\pi\gamma }\alpha(\tau)C_x(0)\right], 
\quad \frac{\d \Pi}{\d t}= \frac{\,\hbarit a\,\alpha'(\tau)}{4\pi m}\,
\int_0^\infty\d \sigma\,\alpha'(\tau-\sigma) C_x(\sigma). \nn
\EEQ
where $\tau=\gamma\, t/2m$ and $C_x$ a function like $f_1$ of Fig. 1a.  
Integrating over the full change
(the whole region where $\alpha'\neq 0$)  one confirms
that the dispersion for a completed, cyclic change 
($\alpha_i=\alpha_f$) of system parameters is nonnegative,
the Thomson formulation of the second law~\cite{ANthomson}. 
It is also nonnegative for noncyclic ($\alpha_i\neq \alpha_f$) 
but completed changes. 

{\it Energy dispersion.}
If $\alpha(\tau)=\alpha_m h(\Omega t)$, with $\Omega$ a  slow 
rate of change, and $\gamma$ is large, expansion in $\Omega$ is 
possible. Omitting numerical factors one gets the structure
\BEA\label{dPIdt=}\frac{\d \Pi}{\d t}=
\,\hbarit\alpha_m^2\Omega^2\,h'\,
\left[\left(\frac{\gamma T}{\,\hbarit a}\right)^2h'
+\frac{\gamma\Omega}{a} h''-\frac{\gamma^2\Omega^2}{a^2}h'''\right]. \EEA
The full integral is always positive and for moderate $T$ the 
first term is dominant. However, for small $T$ the term
$h'h''$ may imply that ${\d \Pi}/{\d t}$ is negative.
This is a firm statement, also valid when starting in equilibrium,
since work is the energy added to the total system. 
It is the more surprising since the relevant domain the characteristic
timescale of the change of $a$,  
$1/\Omega <\,\hbarit^2/(\gamma T^2)$,  may exceed all other timescales,
including the quantum timescale $\,\hbarit/T$, so that applicability
of thermodynamics was to be expected.

{\it Perpetuum mobile with many work extraction cycles.}
The $h'h''$ term can cause extraction of work. 
Let there be $\N$ non-overlapping Gaussians, $h(x)=\exp(-\half x^2)$,
as depicted in Figure 1b. 
Cycles are the intervals $t_i<t<t_{i+1}$ ($i\ge 1$).
Each new cycle is slower than the previous one, $\Omega_{n+1}<\Omega_n$. 
One can make a total number $\N\sim 1/T$ of cycles with equal yield, where
$\N$ is parameterized by $v$, and total yield 
\BEQ \N=\frac{\,\hbarit a}{2\pi\gamma T}\int_{\beta\,\hbarit\Omega_\N/2\pi}
^{\beta\,\hbarit\Omega_1/2\pi}\frac{\d y y}{v+I_1y+I_3y^3},
\quad \W_{\rm tot}=
-\frac{\pi \amsq }{3}\,\frac{\gamma T^2}{\,\hbarit a}
\int_{\beta\,\hbarit\Omega_\N/2\pi}^{
\beta\,\hbarit\Omega_1/2\pi}\d y\frac{v\, y}{v+I_1y+I_3y^3}.
\EEQ 
with $I_1=\half\sqrt{\pi}$, $I_3=3\sqrt{\pi}/4$.
The minus sign of $\W_{\rm tot}$ indicates that
work is performed by the system on the environment. 
This is possible because,
in order to make the extraction cycles, one had to start from the
equilibrium state $\alpha(-\infty)=0$ and change $\alpha$ 
up to $\alpha(t_1)$. In this first part the  
 energy dispersion was 
$\Pi(t_1)=+{\amsq }{\,\hbarit\gamma}\Omega_1^2/24\pi a
\,\, > \,\left|\W_{\rm tot}\right|$.

{\bf Feasibility.} 
The harmonic oscillator can be interpreted as an LC circuit \cite{klim-rev}.
$x$ may correspond to the charge $Q$ on a capacitor, $1/a$ to its
capacitance $C$, $m$ to an inductance $L$, $p$ to a phase $\Phi$,
$\gamma $ to a resistance $R$, and $\eta (t)$ to a random 
electro-motoric force. In this setup there should be
nothing difficult in varying $L\sim m$ or $C\sim 1/a$.
We have proposed to test the violation of the Clausius inequality
in mesocopic linear circuits~\cite{NArlc}. This amounts to measuring
$\langle \delta Q^2\rangle$, $\langle \delta \Phi^2\rangle$ 
and the produced work. The conditions to do this experiment were
reached 20 years ago, and one of the quantities,
$\langle \delta Q^2\rangle$, was measured already and agrees perfectly
with the theoretical predictions ~\cite{w1,flabby}.

{\bf Has thermodynamics been violated?} 
 Let us recall that our results hold also
for $N$ non-interacting Brownian particles in a bath.
Our conclusion is that thermodynamics does not 
always work when, in the quantum regime, one considers 
the Brownian particle in its reduced Hilbert space,
thus  summing out the bath.
We should admit that at low enough $T$ 
the interaction energy between system and bath becomes
relevant, since the damping constant is fixed. 
In a very strict definition of thermodynamics one may
claim that there is no reason why thermodynamics had to apply.

For this reason our surprising findings of the breakdown of the Clausius 
inequality  can in principle be viewed as results outside the domain 
of applicability of thermodynamics.
However, the negativity of the Boltzmann entropy production sets 
in already at moderate $T$, more or less in the same domain where 
energy initially put on the particle starts to dissipate to the bath 
in a non-monotonic fashion, even in limit 
of large damping.

Our second effect in that regime, the presence of many work extraction 
cycles (``perpetuum mobile'') involves only the energy budget of the total
system, and should be more surprising. However, when starting in an
equilibrium state and counting all the energy, the total dispersion
is positive, as demanded by an exact theorem~\cite{ANthomson}.

These extraction cycles relied on the fact that 
the rate of energy dispersion by the total system can be negative.
This is a firm and unexpected statement about the total system,
that may have started in equilibrium.
Thus positivity of the rate of energy dispersion ~\cite{KondPrig} 
is not always a good formulation of the second law.

The new aspects of quantum 
Brownian motion arise from quantum entanglement: A complete description 
in terms of a wave function is possible only for a closed system;
subsystems are necessarily in a mixed state. Thus the quantum Gibbs 
distribution is not an adequate candidate for the description of the
quantum subsystem non-weakly interacting with its thermal bath. 
Connections of these findings with NMR physics~\cite{PAN},
mesoscopic work sources~\cite{ANthomson} and with
Maxwell's demon ~\cite{demon} are discussed elsewhere.

{\it Acknowledgments.}
We thank R. Balian and L.S. Suttorp for discussion.

\end{document}